\documentclass[12pt]{article}
\usepackage{graphicx}
\usepackage{axodraw}
\def\simgt{\,{\rlap{\lower 3.5pt\hbox{$\mathchar\sim$}}\raise 1pt\hbox{$>$}}\,}
\def\simlt{\,{\rlap{\lower 3.5pt\hbox{$\mathchar\sim$}}\raise 1pt\hbox{$<$}}\,}

\def\hph{\hphantom{-}}

\def\gev{~{\rm GeV}}
\def\tev{~{\rm TeV}}

\def\ie{{\it i.e.}}

\newcommand{\beq}{\begin{equation}}
\newcommand{\eeq}{\end{equation}}
\newcommand{\bea}{\begin{eqnarray}}
\newcommand{\eea}{\end{eqnarray}}
\newcommand{\bsub}{\begin{subequations}}
\newcommand{\esub}{\end{subequations} \noindent}

\makeatletter
\newtoks\@stequation

\def\subequations{\refstepcounter{equation}%
  \edef\@savedequation{\the\c@equation}%
  \@stequation=\expandafter{\theequation}
  \edef\@savedtheequation{\the\@stequation}
  \edef\oldtheequation{\theequation}%
  \setcounter{equation}{0}%
  \def\theequation{\oldtheequation\alph{equation}}}

\def\endsubequations{%
  \ifnum\c@equation < 2 \@warning{Only \the\c@equation\space subequation
    used in equation \@savedequation}\fi
  \setcounter{equation}{\@savedequation}%
  \@stequation=\expandafter{\@savedtheequation}%
  \edef\theequation{\the\@stequation}%
  \global\@ignoretrue}

\def\eqnarray{\stepcounter{equation}\let\@currentlabel\theequation
\global\@eqnswtrue\m@th
\global\@eqcnt\z@\tabskip\@centering\let\\\@eqncr
$$\halign to\displaywidth\bgroup\@eqnsel\hskip\@centering
     $\displaystyle\tabskip\z@{##}$&\global\@eqcnt\@ne
      \hfil$\;{##}\;$\hfil
     &\global\@eqcnt\tw@ $\displaystyle\tabskip\z@{##}$\hfil
   \tabskip\@centering&\llap{##}\tabskip\z@\cr}

\makeatother

\begin{document}
\thispagestyle{empty}
\vspace*{-15mm}
\baselineskip 10pt
\begin{flushright}
\begin{tabular}{l}
{\bf OCHA-PP-183}\\
{\bf KEK-TH-792}\\
{\bf hep-ph/0112163}
\end{tabular}
\end{flushright}
\baselineskip 24pt 
\vglue 10mm 
\begin{center}
{\Large\bf
The stau exchange contribution to muon $g-2$ in the decoupling 
solution
}
\\
\vspace{8mm}

\baselineskip 18pt 
\def\thefootnote{\fnsymbol{footnote}}
\setcounter{footnote}{0}

{\bf Gi-Chol Cho$^{1)}$, Naoyuki Haba$^{2)}$ and Junji Hisano$^{3)}$}
\vspace{5mm}

$^{1)}${\it 
Department of Physics, Ochanomizu University, Tokyo 112-8610, Japan}\\
$^{2)}${\it 
Faculty of Engineering, Mie University, Tsu, Mie, 514-8507, Japan}\\
$^{3)}${\it 
Theory Group, KEK, Oho 1-1, Tsukuba, Ibaraki 305-0801, Japan}
\vspace{10mm}
\end{center}
\begin{center}
{\bf Abstract}\\[7mm]
\begin{minipage}{14cm}
\baselineskip 16pt
\noindent
We study the possibility that the lepton-flavor changing process 
can induce the suitable magnitude of the muon anomalous magnetic 
moment ($g_\mu -2$) in the decoupling solution to the flavor problem 
in the minimal supersymmetric standard model. 
Our analyses introduce the flavor mixings of left- and right-handed 
stau and smuon phenomenologically.   
It is found that if both the left- and right-handed sleptons 
have sizable flavor mixings, the correction to $g_\mu -2$ from the 
lighter slepton can reach to 10$^{-9}$ while the correction to 
the branching ratio of $\tau \to  \mu \gamma$ satisfies the current 
experimental bound. 
On the other hand, when only the left-handed or right-handed 
sleptons have the large flavor mixing, the suitable magnitude of 
the correction to $g_\mu-2$ is not realized owing to the experimental 
bound of $\tau\to \mu \gamma$. 
\end{minipage}
\end{center}
\newpage
\baselineskip 18pt 
While the minimal supersymmetric (SUSY) standard model (MSSM) is 
one of the promising candidates beyond the standard model (SM), 
introduction of the SUSY breaking terms may lead to the flavor 
changing neutral current (FCNC) and CP problems. 
The decoupling solution (sometime called as ``effective 
SUSY'')~\cite{effsusy} has been proposed to solve these problems. 
In this solution, the squarks and sleptons in the first and second 
generations are heavy enough so that their contributions to the 
FCNC or CP violating processes are sufficiently suppressed.  
On the other hand, the gauginos, higgsinos and the sfermions in 
the third generation are appropriately light to satisfy the 
naturalness condition on the Higgs boson mass. 
In other words, the sfermions whose fermionic partners
have the large Yukawa couplings belong to the ``lighter group'' 
because they are responsible for the quantum corrections to the 
Higgs boson mass. 

The muon anomalous magnetic moment, conventionally parameterized 
as $a_\mu \equiv (g_\mu -2)/2$, has been measured precisely 
by the E821 experiment at the Brookhaven National Laboratory. 
The current world average of $a_\mu$ is about 2.6-$\sigma$ away 
from the SM prediction~\cite{Brown:2001mg}: 
\begin{eqnarray}
a_\mu({\rm exp})-a_\mu({\rm SM})=426(165) \times 10^{-11}, 
\label{g_2_exp}
\end{eqnarray}
which may suggest that new physics exists around the TeV 
scale~\cite{Czarnecki:2001pv}. 
After the announcement, the many works to understand the 
implication of (1) to physics beyond the SM have been 
done~\cite{g-2_recent}.  
It will take more time to confirm if (1) is really a signal 
of new physics from theoretical and experimental points of view. 
However, once it is confirmed, the constraint on models beyond 
the SM will be stronger. 

It has been considered that the decoupling solution is disfavored 
from this observation, since the smuon and the muon-sneutrino are 
too heavy to give a sizable contribution to $g_\mu-2$. 
However, if the stau and the tau-sneutrino have the couplings to 
the muon in this model, their mediation may be able to enhance 
$g_\mu-2$ suitably. 
When $\tan\beta$ is small, the Yukawa coupling of tau lepton is 
also small, and it is model dependent whether the stau is classified 
to the lighter or heavier group in the decoupling solution.  
In the case, even if one slepton belongs to the lighter group, 
the slepton may not necessarily correspond to the scalar 
component of the tau-lepton superfield and may be some mixture 
of the stau and the smuon. 
The neutrino oscillation data~\cite{atm} may suggest the 
left-handed stau and smuon have a large mixing~\cite{atmano}. 
If the energy scale of the SUSY-breaking mediation to the MSSM 
is higher than the mass scale of the right-handed neutrino,  
this spectrum should be realized so that the tau-neutrino 
Yukawa coupling does not destabilize the naturalness of the Higgs 
boson mass.

In this paper, we examine the possibility that the lepton-flavor 
changing process can induce the suitable magnitude of the muon 
anomalous magnetic moment in the decoupling solution to the 
flavor problem. 
We introduce the mixings of the left- and right-handed stau and 
smuon phenomenologically. 
We find that if both the left- and right-handed sleptons have 
the sizable mixings between the second and the third generations, 
the correction to $g_\mu -2$ from the lighter slepton can reach to 
$10^{-9}$ which corresponds to about 2-$\sigma$ lower bound 
of $a_\mu$, 
while the correction to the branching ratio of $\tau\rightarrow 
\mu \gamma$ satisfies the current experimental bound. 
When only the left-handed sleptons have a large mixing between 
the second and third generations, the supersymmetric contribution 
to $g_\mu -2$ is constrained to be smaller than $10^{-10}$ from 
the experimental bound of $\tau\rightarrow \mu \gamma$. 
On the other hand, in the case with a large mixing only 
 in the right-handed slepton between the second and third 
 generations,  the magnitude of the model contribution 
to $g_\mu -2$ is at most a few $10^{-10}$, and the contribution 
is positive when the higgsino mass (the $\mu$-term) is negative.

Let us briefly review the muon anomalous magnetic moment and  
the radiative decay $\tau \to \mu \gamma$ in the SUSY-SM with 
the lepton flavor violations (LFV). 
The effective Lagrangian related to $g_\mu-2$ and 
$\tau \to \mu \gamma$ is as follows: 
\begin{eqnarray}
{\cal L}_{eff} = e \frac{m_{\ell_j}}{2}\bar{\ell}_i \sigma_{\mu\nu}  
F^{\mu\nu} (L_{ij} P_L+R_{ij} P_R) \ell_j, 
\label{eff_op}
\end{eqnarray}
where $m_{\ell_i}$ is a mass of the charged lepton $\ell_i$ 
and the indices $i,j=1,2,3$ denote the generation. 
The coefficients $L_{ij}$ and $R_{ij}$ account for 
the model dependent contributions to the dipole operators 
in the quantum level. 
Then, the muon anomalous magnetic moment is given in terms of 
$L_{ij}$ and $R_{ij}$ as 
\begin{eqnarray}
a_\mu = m_\mu^2 (L_{22}+R_{22}), 
\label{eq:amu}
\end{eqnarray}
while the branching ratio of $\tau \to \mu \gamma$ is 
given by 
\begin{eqnarray}
{\rm Br}(\tau \to \mu \gamma) 
= 
{\rm Br}(\tau \to \mu \nu_\tau\bar{\nu}_\mu)~\frac{48\pi^3 \alpha}
{G_F^2 } (|L_{23}|^2+|R_{23}|^2) ,
\label{eq:branc}
\end{eqnarray}
where $G_F$ is the Fermi constant and 
${\rm Br}(\tau \to \mu \nu_\tau\bar{\nu}_\mu) =17\%$.
In the SUSY-SM with LFV, $L_{ij}$ and $R_{ij}$ are given by 
contributions from (i) chargino-sneutrino exchange, and 
(ii) neutralino-charged slepton exchange.  
The explicit form of $L_{ij}$ and $R_{ij}$ in the framework 
of the SUSY-SM with LFV can be found, for example, in 
ref.~\cite{Hisano:1996cp}. 

In the effective SUSY-SM, the supersymmetric contributions to 
both $g_\mu -2$ and $\tau \to \mu \gamma$ may be dominated by the 
stau or the tau-sneutrino exchange through the lepton flavor 
violating interactions.\footnote{
Note that it is aruged in Ref.~\cite{Chen:2001kn} that two-loop
diagrams from an anomalous $H^--W^+-\gamma$ coupling may be a sizable
contribution to $g_\mu -2$ for large $\tan\beta$, large $\mu$
parameter, and light stop and sbottom. }
The origin of the flavor violating interactions is the flavor 
off-diagonal terms in the soft SUSY breaking mass matrix for 
the sleptons. In the following analysis, we focus on the mixing between 
the second and third generations of sleptons. 
In the limit that the left-right mixing terms of the charged 
sleptons are ignored, the mass eigenstates are given using 
the mixing angles $\theta_L$ and $\theta_R$ as follows: 
\begin{eqnarray}
\left(
\begin{array}{c}
\tilde{\mu}_L'\\
\tilde{\tau}_L'
\end{array}
\right)
&=&
\left(
\begin{array}{cc}
\cos\theta_L&
-\sin\theta_L\\
\sin\theta_L&
\hph \cos\theta_L
\end{array}
\right)
\left(
\begin{array}{c}
\tilde{\mu}_L\\
\tilde{\tau}_L
\end{array}
\right)
\,,
\label{leftmixing}
\\
\left(
\begin{array}{c}
\tilde{\mu}_{R}'\\
\tilde{\tau}_{R}'
\end{array}
\right)
&=&
\left(
\begin{array}{cc}
\cos\theta_R&
-\sin\theta_R\\
\sin\theta_R&
\hph 
\cos\theta_R
\end{array}
\right)
\left(
\begin{array}{c}
\tilde{\mu}_R\\
\tilde{\tau}_R 
\end{array}
\right), 
\end{eqnarray}
where $\widetilde{\mu}_{L(R)}$ and $\widetilde{\tau}_{L(R)}$ are the
flavor eigenstates for left(right)-handed sleptons. The mass
eigenstates $\widetilde{\mu}'_{L(R)}$ are heavier than
$\widetilde{\tau}'_{L(R)}$. The flavor mixing of the sneutrinos
between the second and the third generations are describe by the same
unitary matrix with the charged sleptons (r.h.s. in
Eq.~(\ref{leftmixing})).

The mass eigenvalues and the mixings are related to the SUSY breaking
parameters in the Lagrangian as
\begin{eqnarray}
m_{\tilde{\tau}_{L/R}}^2
&=&\frac12 
\left\{(m^2_{L/R})_{33}+(m^2_{L/R})_{22}
\right.
\nonumber\\
&&\left.
-\sqrt{
((m^2_{L/R})_{33}-(m^2_{L/R})_{22})^2
+4|(m^2_{L/R})_{12}|^2}
\right\},
\nonumber\\
\tan2\theta_{L/R}
&=&
\frac{2(m^2_{L/R})_{12}}{(m^2_{L/R})_{22}-(m^2_{L/R})_{33}}.
\end{eqnarray}
where $(m^2_{L/R})_{ij}$ are for the SUSY breaking parameters of the
left-handed (right-handed) sleptons and $i,j$ refer to the generation.
Thus, in order to generate lighter stau with the the large mixing, 
$(m^2_{L/R})_{33}\simeq (m^2_{L/R})_{22}\simeq (m^2_{L/R})_{23}$ is 
required. 


In a limit where the relevant SUSY breaking parameters are 
given by a common mass $m_{\rm SUSY}$, the SUSY contribution 
to the muon $g_\mu-2 (a_\mu^{\rm SUSY})$ is approximately 
given as 
\begin{eqnarray}
a_\mu^{\rm SUSY} &\simeq&
\frac{5\alpha_2+\alpha_Y}{48\pi}
\frac{m_\mu^2\tan\beta}{m^2_{\rm SUSY}} 
\sin^2\theta_L
-
\frac{\alpha_Y}{24\pi}
\frac{m_\mu^2 \tan\beta }{m^2_{\rm SUSY}} 
\sin^2\theta_R
\nonumber\\
&& +
\frac{\alpha_Y}{96\pi}
\frac{m_\mu m_\tau \tan\beta }{m^2_{\rm SUSY}} 
\sin2\theta_L \sin2\theta_R.
\label{eq:amu:approx}
\end{eqnarray}
Here, we keep terms proportional to $\tan\beta$. 
On the other hand, coefficients $L_{23}$ and $R_{23}$ 
in Eq.~(\ref{eff_op}) which give $\tau \to \mu \gamma$ 
(\ref{eq:branc}) are expressed as 
\begin{eqnarray}
L_{23} 
&\simeq& -
\frac{\alpha_Y}{192\pi}\frac{\tan\beta}{m^2_{\rm SUSY}}
\sin2\theta_R
+ 
\frac{\alpha_Y}{198\pi}\frac{\tan\beta}{m^2_{\rm SUSY}}
\sin2\theta_R
\cos^2\theta_L
\,,
\label{eq:l23}
\\ 
R_{23} 
&\simeq& 
\hph 
\frac{5\alpha_2+\alpha_Y}{384\pi}\frac{\tan\beta}{m^2_{\rm SUSY}}
\sin2\theta_L
+ 
\frac{\alpha_Y}{198\pi}\frac{\tan\beta}{m^2_{\rm SUSY}}
\sin2\theta_L
\cos^2\theta_R
\,.
\label{eq:r23}
\end{eqnarray}
The SUSY contributions, Eqs.~(\ref{eq:amu:approx}), (\ref{eq:l23}) 
and (\ref{eq:r23}), can be calculated from four diagrams in 
Fig.~1(a)-1(d). 
Fig.~1(a) gives the first terms in Eqs.~(\ref{eq:amu:approx}) 
and (\ref{eq:r23}), while Fig.~1(b) gives 
the second term in Eq.~(\ref{eq:amu:approx}) and the first term 
in Eq.~(\ref{eq:l23}). 
The last terms in Eqs.~(\ref{eq:amu:approx})-(\ref{eq:r23}) 
correspond to Figs.~1(c) and 1(d). 
In the above, we retain the terms which are dominant when 
$\tan\beta$ is small ($\simlt 10$ but larger than one). 
They are useful to understand the following numerical result.  

It is worth to note that each diagram in Fig.~1 is proportional 
to $M_1\mu$ or $M_2\mu$, where $M_1$ and $M_2$ are the U(1)$_Y$ 
and SU(2)$_L$ gaugino masses, respectively, and $\mu$ denotes 
the higgsino mass. 
This proportionality is not explicitly shown in 
Eqs.~(\ref{eq:amu:approx})-(\ref{eq:r23}) because the equations 
are derived in the limit where all SUSY particles have the common 
mass $m_{\rm SUSY}$ so that $M_1=M_2=\mu=m_{\rm SUSY}$ 
are cancelled by some powers of $m_{\rm SUSY}$ in the denominator 
which come from the propagators. 
When only the right-handed sleptons have the flavor mixing, 
the contribution to $g_\mu-2$ tends to be positive for $M_1\mu<0$, 
that is favored from the current measurement of muon 
$g-2$ (\ref{g_2_exp}). 
In the following numerical study, $\mu$ is taken to be negative 
when only the right-handed sleptons have the flavor mixing, while 
it is positive for the other cases. 
\begin{figure}[t]
\begin{center} 
\begin{picture}(100,250)(260,50)
\ArrowLine(100,150)(160,150)
\Text(130,145)[t]{$\mu_R/\tau_R$}
\CArc(210,150)(50,0,180) 
\Vertex(210,200){3}
\Text(150,190)[c]{${\widetilde H}^\pm$(${\widetilde H}^0$)}
\Text(280,190)[c]
     {${\widetilde W}^\pm$(${\widetilde W}^0$,${\widetilde B}^0$)}
\DashArrowLine(260,150)(160,150){5}
\Text(210,145)[t]{${\widetilde \nu}_{\tau}'$(${\widetilde \tau}_{L}'$)}
\ArrowLine(320,150)(260,150)      \Text(290,145)[t]{$\mu_L$}
\Photon(245,125)(285,100){4}{4}   \Text(290,95)[lt]{$\gamma$}
\Text(210,80)[t]{{\Large{(a)}}}
\ArrowLine(420,150)(360,150)  
\Text(390,145)[t]{$\mu_L/\tau_L$}    
\CArc(470,150)(50,0,180)
\Vertex(470,200){3} 
\Text(420,190)[c]{${\widetilde H}^0$}
\Text(520,190)[c]{${\widetilde B}^0$}
\DashArrowLine(420,150)(520,150){5}
\Text(470,145)[t]{${\widetilde \tau}'_{R}$}
\ArrowLine(520,150)(580,150)  \Text(550,145)[t]{$\mu_R$}
\Photon(530,100)(485,125){4}{4}   \Text(535,95)[lt]{$\gamma$}
\Text(470,80)[t]{{\Large{(b)}}}
\end{picture}
\end{center}
\vspace{-3cm}

\begin{center} 
\begin{picture}(100,250)(260,50)
\ArrowLine(100,150)(160,150)
\Text(130,145)[t]{$\mu_R/\tau_R$}
\CArc(210,150)(50,0,180) 
\Text(210,190)[c]{${\widetilde B}^0$}
\DashArrowLine(260,150)(210,150){5}
\DashArrowLine(160,150)(210,150){5}
\Vertex(210,150){3}
\Text(180,145)[t]{${\widetilde \tau}'_{R}$}
\Text(240,145)[t]{${\widetilde \tau}'_{L}$}
\ArrowLine(320,150)(260,150)      \Text(290,145)[t]{$\mu_L$}
\Photon(245,125)(285,100){4}{4}   \Text(290,95)[lt]{$\gamma$}
\Text(210,80)[t]{{\Large{(c)}}}
\ArrowLine(420,150)(360,150)
\Text(390,145)[t]{$\mu_L/\tau_L$}    
\CArc(470,150)(50,0,180)
\Text(470,190)[c]{${\widetilde B}^0$}
\Vertex(470,150){3}
\DashArrowLine(470,150)(520,150){5}
\DashArrowLine(470,150)(420,150){5}
\Text(440,145)[t]{${\widetilde \tau}_{L}'$}
\Text(500,145)[t]{${\widetilde \tau}_{R}'$}
\ArrowLine(520,150)(580,150)  \Text(550,145)[t]{$\mu_R$}
\Photon(530,100)(485,125){4}{4}   \Text(535,95)[lt]{$\gamma$}
\Text(470,80)[t]{{\Large{(d)}}}
\end{picture}
\end{center}
\caption{Feynman diagrams of $g_\mu-2$ and $\tau\rightarrow\mu\gamma$, 
which are proportional to $\tan\beta$. 
The symbols $\widetilde{B}^0$, $\widetilde{W}^0$ ($\widetilde{W}^\pm$), 
and $\widetilde{H}^0$ ($\widetilde{H}^\pm$) are bino, neutral 
(charged) wino, and neutral (charged) higgsino, respectively. 
The lightest left-handed (right-handed) charged slepton is denoted 
by $\widetilde{\tau}_{L}'$ ($\widetilde{\tau}_{R}'$), 
while $\widetilde{\nu}_{\tau}'$ being the sneutrino.}
\label{fig1:figures}
\end{figure}
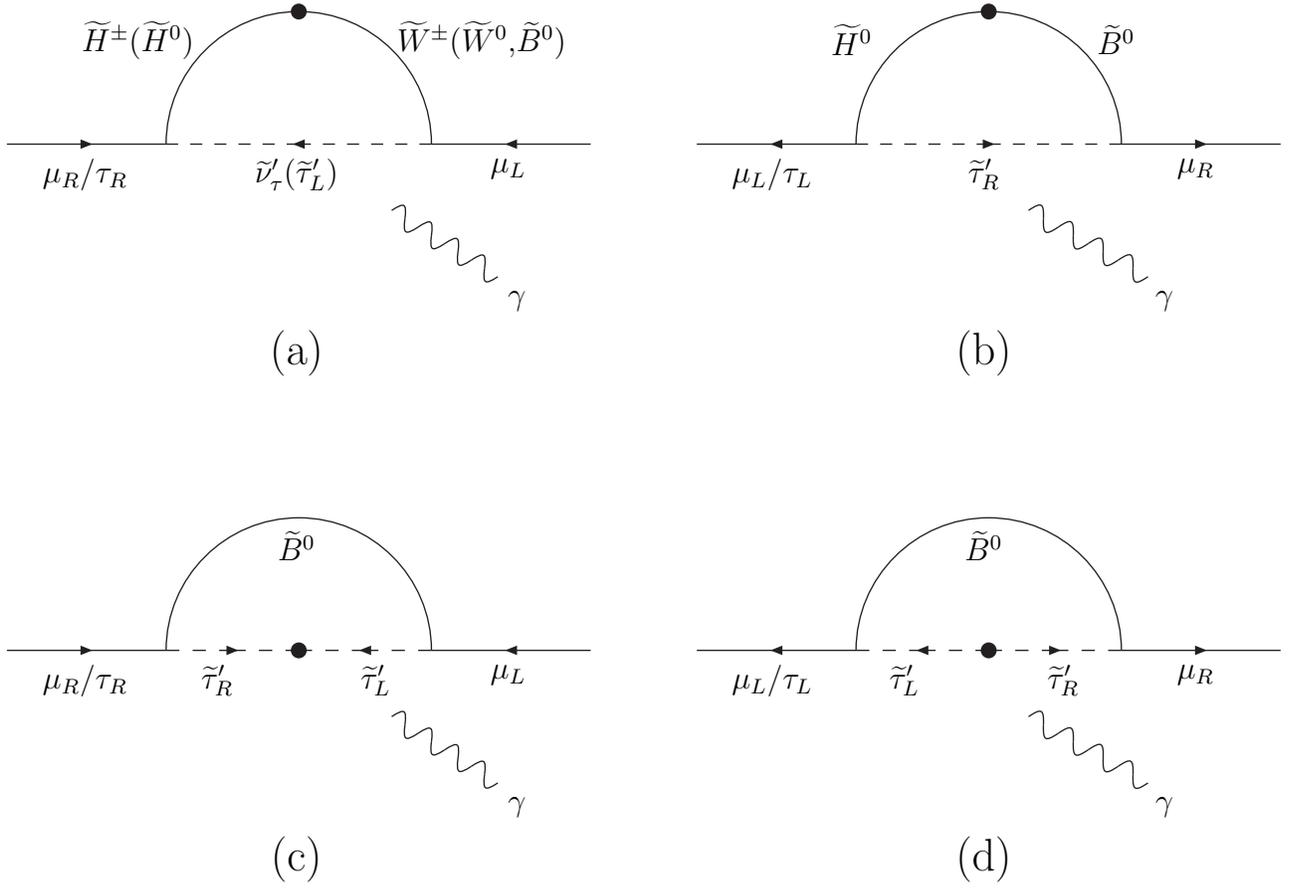

\begin{figure}[t]
\begin{center}
\includegraphics[width=10cm,clip]{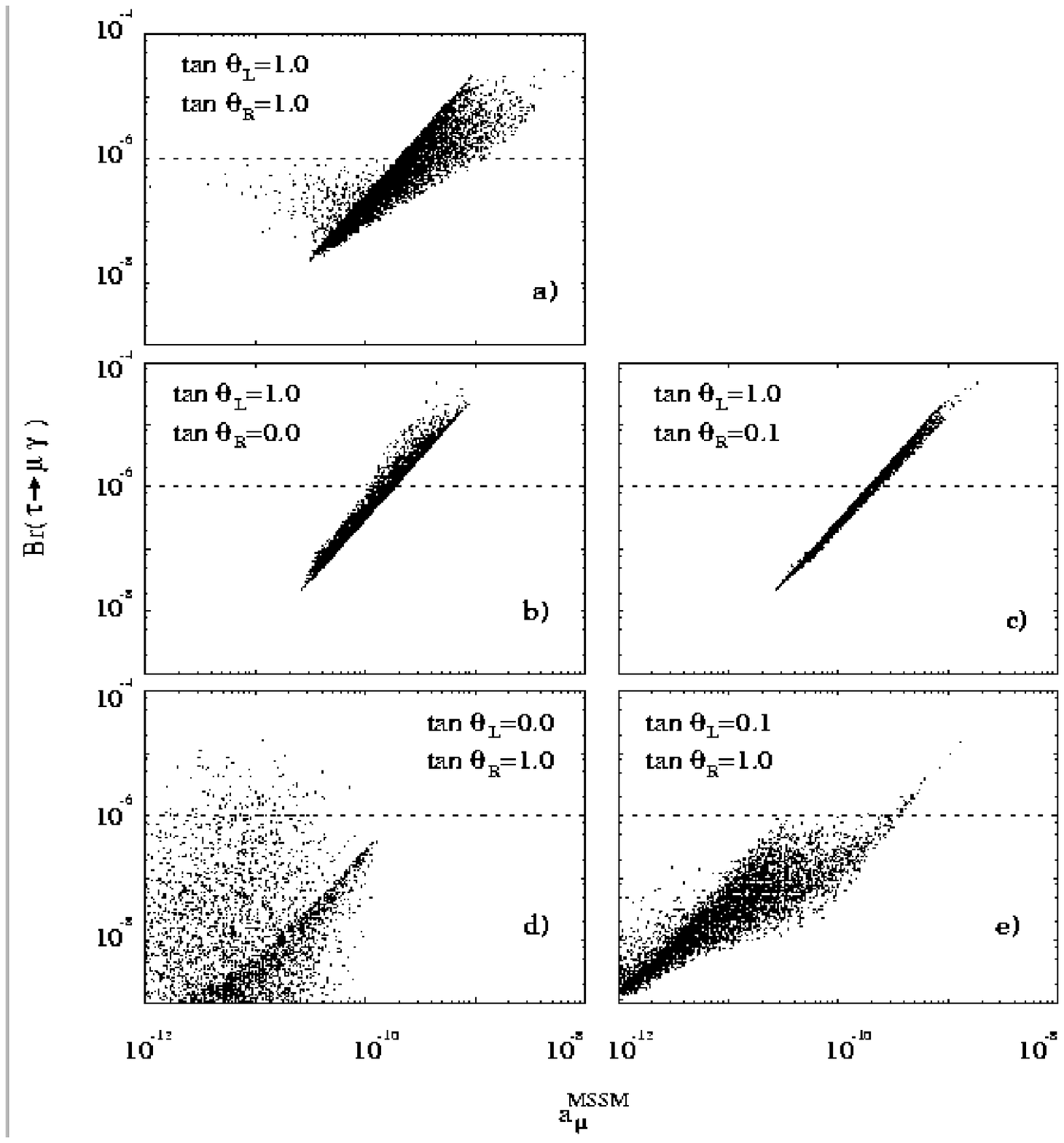}
\end{center}
\caption{
Prediction of the effective SUSY-SM on the muon $g-2$ 
and the branching ratio $\tau \to \mu \gamma$ for $\tan\beta=3$. 
The dependence of the mixing angles $\theta_L$ and $\theta_R$ 
are examined for $(\tan\theta_L,\tan\theta_R) = (1.0,1.0)$ (a), 
$(1.0,0.0)$ (b), $(1.0,0.1)$ (c), $(0.0,1.0)$ (d) and 
$(0.1,1.0)$ (e). 
The soft SUSY breaking masses for the SU(2)$_L$ doublet and 
singlet sleptons in the third generation are taken as 
$100\gev-1000\gev$ while those in the second generation are 
fixed by $10\tev$. 
The higgsino mass and the SU(2)$_L$ gaugino mass are 
examined between $100\gev$ and $500\gev$, and the $A$-terms for 
the charged sleptons are fixed by zero. 
The relative sign between the higgsino and the gaugino masses 
is taken to be positive for (a)-(c) and (e), while it is negative 
for (d). 
The bounds on the stau and chargino from the direct search 
experiment, $m_{\widetilde{\tau}} >85\gev$ and 
$m_{\widetilde{\chi}^-_1} > 103.5\gev$, are taken into 
account~\cite{sparticles:limit}. 
The horizontal dotted-line in each figure denotes the 90\% CL 
upper bound of ${\rm Br}(\tau \to \mu \gamma)$~\cite{belltau}. 
}
\label{fig:figures}
\end{figure}
Let us examine the muon $g-2$ and the branching ratio of 
$\tau \to \mu \gamma$ in the effective SUSY-SM quantitatively, 
taking account of the flavor mixing of sleptons between the 
second and third generations. 
We show the prediction of the effective SUSY-SM on the muon $g-2$ 
and ${\rm Br}(\tau \to \mu \gamma)$ for $\tan\beta = 3$ 
in Fig.~\ref{fig:figures}. 
The dependence of the mixing angles $\theta_L$ and $\theta_R$ are 
examined for $(\tan\theta_L,\tan\theta_R) = (1.0,1.0)$ (a), 
$(1.0,0.0)$ (b), $(1.0,0.1)$ (c), $(0.0,1.0)$ (d) and 
$(0.1,1.0)$ (e). 
The soft SUSY breaking masses for the SU(2)$_L$ doublet and 
singlet sleptons in the third generation are taken as 
$100-1000\gev$ while those in the second generation are fixed 
by $10\tev$. 
The higgsino mass $\mu$ and the SU(2)$_L$ gaugino mass $M_2$ are 
taken as $100\gev \le |\mu|, M_2 \le 500\gev$ with the GUT relation 
$M_2/M_1 = 3 \alpha_2/5 \alpha_Y$. 
The $A$-terms for the charged sleptons are fixed at zero. 
The relative sign between the higgsino and the gaugino masses 
is taken to be positive for (a)-(c) and (e), 
while it is negative for (d), as mentioned above. 
In the analysis, the bounds on the stau and the chargino 
masses from the direct search experiment, 
$m_{\widetilde{\tau}} >85\gev$ and 
$m_{\widetilde{\chi}^-_1} > 103.5\gev$, are taken into 
account~\cite{sparticles:limit}. 
The horizontal dotted-line in each figure shows the experimental 
bound on ${\rm Br}(\tau \to \mu \gamma)$ at 90\% CL~\cite{belltau},  
\begin{equation}
{\rm Br}(\tau \to \mu \gamma) < 1.0 \times 10^{-6}. 
\label{br.exp}
\end{equation}

Let us first see Fig.~\ref{fig:figures}(a). 
Here, we examine the case in which the flavor mixing angles 
for both the left- and the right-handed sleptons are assumed 
to be maximal, \ie, $\theta_L=\theta_R=\pi/4$. 
It seems that there may be a certain proportionality between 
$a_\mu^{\rm SUSY}$ and $\tau \to \mu \gamma$. 
This is because both $a_\mu^{\rm SUSY}$ and 
$\tau \to \mu \gamma$ are given by the coefficients 
$L_{ij}$ and $R_{ij}$ as Eqs.~(\ref{eq:amu}) and 
(\ref{eq:branc}). 
Fig.~\ref{fig:figures}(a) shows that $a_\mu^{\rm SUSY}$ is 
predicted to be much larger than that in the other four 
figures. 
When the flavor mixings for both the left- and right-handed 
sleptons are sizable and their magnitudes are close to each 
other, the last term in Eq.~(\ref{eq:amu:approx}) may become 
sizable because the term is proportional to $m_\tau$ due to 
the left-right mixing of the tau-sleptons. 
Since this effect disappears when one of the mixing angles  
is much smaller than the others, this is why $a_\mu^{\rm SUSY}$ 
could be large comparing with other four sets of the 
mixing angles, Figs.~\ref{fig:figures}(b)-(e). 
Taking account of the bounds on ${\rm Br}(\tau \to \mu \gamma)$ 
(\ref{br.exp}), we find that $a_\mu^{\rm SUSY} \sim 10^{-9}$, 
which corresponds to about 2-$\sigma$ lower bound, 
is expected in the certain region of the parameter space. 

We show $a_\mu^{\rm SUSY}$ and $\tau \to \mu\gamma$ 
when the flavor mixing of the left-handed sleptons is 
dominantly large $(\tan\theta_L=1.0, \tan\theta_R \sim 0)$ 
in Fig.~\ref{fig:figures}(b) and (c). 
In the small $\theta_R$ limit, the first term in 
Eq.~(\ref{eq:amu:approx}) and $R_{23}$ in Eq.~(\ref{eq:r23}) 
dominate $a_\mu^{\rm SUSY}$ and $\tau \to \mu \gamma$, 
respectively, while 
the other two terms in Eq.~(\ref{eq:amu:approx}) and $L_{23}$ 
are highly suppressed by small $\theta_R$. 
We find in Fig.~\ref{fig:figures}(b) and (c) 
that $a_\mu^{\rm SUSY}$ and $\tau \to \mu \gamma$ are 
strongly correlated each other and it may be difficult to 
obtain the suitable magnitude of the muon $g-2$ in this set 
of the mixing angles by taking account of the experimental bound 
of $\tau \to \mu \gamma$. 

Figs.~\ref{fig:figures}(d) and (e) show $a_\mu^{\rm SUSY}$ and 
$\tau \to \mu \gamma$ 
between the second and third generations are maximal for the 
right-handed sleptons ($\tan\theta_R=1.0$), but very small or 
zero for the left-handed sleptons ($\tan\theta_L=0.1$ or 0). 
We find that this set of the mixing angles makes both 
$a_\mu^{\rm SUSY}$ and $\tau \to \mu \gamma$ to be smaller than 
the other set of the mixing angles. 
As seen in (\ref{eq:amu:approx}), $a_\mu^{\rm SUSY}$ is proportional 
to the U(1)$_Y$ gauge coupling because the first and third terms 
in (\ref{eq:amu:approx}) are suppressed due to small $\sin\theta_L$. 
On the other hand, only the coefficient $L_{23}$ (\ref{eq:l23}) 
is responsible for $\tau \to \mu\gamma$, since $R_{23}$ (\ref{eq:r23}) 
are negligible in the small $\theta_L$ limit. 
Furthermore, $L_{23}$ itself also becomes small because both terms 
in Eq.~(\ref{eq:l23}) may be cancelled out for small $\theta_L$. 
Then, we find that almost all parameter region satisfies 
the bound from $\tau \to \mu\gamma$, but suitable magnitude 
of $a_\mu^{\rm SUSY}$ cannot be expected in the region. 

We have so far studied $a_\mu^{\rm SUSY}$ and 
$\tau \to \mu\gamma$ in the effective SUSY-SM quantitatively.  
Our results are restricted in the case of $\tan\beta=3$. 
One may have an interest for the large $\tan\beta$ case. 
Let us recall the expressions of $a_\mu^{\rm SUSY}$ and 
$\tau \to \mu\gamma$ in the common SUSY mass limit, 
Eqs.~(\ref{eq:amu:approx})-(\ref{eq:r23}).  
The $\tan\beta$ dependence of $a_\mu^{\rm SUSY}$ is linear 
while that of ${\rm Br}(\tau \to \mu\gamma)$ is quadratic 
because $|L_{23}|^2+|R_{23}|^2$ appears in the branching ratio. 
Thus, the sizable enhancement of $a_\mu^{\rm SUSY}$ may be 
possible for large $\tan\beta$, but ${\rm Br}(\tau \to \mu\gamma)$ 
is also much enhanced by $\tan^2\beta$ 
so that it is disfavored from the current experimental bound. 

The numerical study in the above tells us that, in order to explain
simultaneously the experimental data of the muon $g-2$ and the
radiative decay $\tau \to \mu \gamma$ in the framework of the
effective SUSY-SM, not only the flavor mixing between the second and
third generations of the left-handed sleptons, but also that of the
right-handed sleptons must be large. In addition, the mixing between
the left- and right-handed tau-sleptons plays an important role to
induce the suitable magnitude of the muon $g-2$.
It is worth to look for the model of effective SUSY where the flavor
mixing of both the left- and right-handed sleptons are predicted to be
large, but it is beyond our scope in this paper.

In summary,  we have examined the possibility that the 
lepton-flavor changing process can induce the suitable magnitude 
of the muon anomalous magnetic moment in the effective SUSY-SM. 
In our analysis, the slepton flavor mixings between the second and 
third generations are introduced phenomenologically. 
We find that if both the left- and right-handed sleptons have 
sizable flavor mixings, the supersymmetric contributions to 
$g_\mu -2$ from the lighter slepton can reach to 10$^{-9}$ which 
corresponds to about 2-$\sigma$ lower bound of $a_\mu$, while 
the correction to the branching ratio of $\tau\to \mu \gamma$ 
satisfies the current experimental bound. 
The constraints from $g_\mu-2$ and $\tau \to \mu\gamma$ cannot 
be satisfied simultaneously when there is the lack of the flavor 
mixing in either the left- or the right-handed sleptons.  
When only the left-handed slepton have the large flavor mixing 
between the second and the third generations, 
the supersymmetric contribution to $g_\mu -2$ is constrained 
to be smaller than $10^{-10}$ owing to the experimental bound 
of $\tau\rightarrow \mu \gamma$. 
We also find that the suitable magnitude of $a_\mu^{\rm SUSY}$ 
cannot be expected when only the right-handed sleptons have 
the large flavor mixing although almost all parameter region 
satisfies the bound from $\tau \to \mu\gamma$.  

\section*{Note added}
After completion of this paper, we found that the estimate 
of the hadronic light-by-light scattering was 
revisited~\cite{Knecht:2001qf}. 
The new estimate~\cite{Knecht:2001qf} tells us that the sign 
of pseudo-scalar pole contribution is opposite from the 
previous estimates \cite{Bijnens:1995xf} and it was confirmed 
in refs.~\cite{Hayakawa:2001bb,czarnecki} recently.  
Using the new value of the light-by-light scattering contribution 
in \cite{Hayakawa:2001bb}, we find that Eq.~(\ref{g_2_exp}) 
becomes $a_\mu({\rm exp})-a_\mu({\rm SM})=252(164) \times 10^{-11}$, 
and the discrepancy between the experimental value and the SM 
prediction is 1.5-$\sigma$. 
Our study in this paper is valid even under this reduction of 
the deviation. 

\section*{Acknowledgment}
This work is supported in part by the Grant-in-Aid for Science 
Research, Ministry of Education, Science and Culture, Japan 
(Priority Area 707 `Supersymmetry and Unified Theory of Elementary 
Particles', No. 13001292 for J.H., No.12740146 for N.H., 
and No.13740149 for G.C.C.).


\end{document}